\documentclass[conference]{IEEEtran}
\IEEEoverridecommandlockouts

\usepackage{amsfonts}
\usepackage{dsfont} 
\usepackage{setspace}
\usepackage{color}
\usepackage{amssymb}
\usepackage{cite}
\usepackage[cmex10]{amsmath}
\usepackage{algorithm}
\usepackage{algorithmic} 
\usepackage{array}
\usepackage{mathrsfs}
\usepackage{graphicx}
\usepackage{latexsym}
\usepackage{amscd}
\usepackage{amsfonts}
\usepackage{subfigure}
\usepackage{amsmath,amscd,amssymb,verbatim}
\usepackage[utf8]{inputenc}
\usepackage{graphics}
\usepackage{amsthm}
\usepackage[T1]{fontenc}
\usepackage[utf8]{inputenc}
\usepackage{authblk}
\usepackage[nocomma]{optidef}
\usepackage{mathtools}
\usepackage{makecell}
\usepackage{multirow}
\usepackage{anyfontsize} 
\usepackage{soul}
\usepackage{diagbox}
\usepackage{textcomp}
\usepackage{xcolor}
\usepackage{blindtext}
\IEEEoverridecommandlockouts
\usepackage[top=.7in, bottom=1.0in, left=.65in, right=.65in]{geometry}

\usepackage{color}

\begin{document}
\title{Minimizing Quantized Semantic Age of Information (QSAoI) in Foundation Model-Based Semantic Communications\vspace{-0.3cm}}
\author{
	Huanyu Zhang$^{\dag}$, Yulin Hu$^{\ddag}$, Xiaopeng Yuan$^{\dag}$, Aydin Sezgin$^{\S}$, and Anke Schmeink$^\dag$  \\ 
    $^\dag$INDA Chair, RWTH Aachen University, Germany, Email: $zhang|yuan|schmeink$@inda.rwth-aachen.de \\
	$^\ddag$School of Electronic Information, Wuhan University, China, 
	Email: $yulin.hu$@whu.edu.cn \\
    $^{\S}$Department of Digital Communication Systems, Ruhr University Bochum, Germany, Email: $aydin.sezgin$@rub.de 
\vspace{-0.6cm}
}
\maketitle
 

\begin{abstract}

The emerging techniques of semantic communications and edge computing in 6G networks necessitate a paradigm shift toward co-designed semantic-aware and adaptive resource allocation for short-packet transmissions.
However, there is a fundamental gap between the semantic layer and the physical layer under low-latency finite blocklength (FBL) effects.
To bridge this gap, we introduce the Quantized Semantic Age of Information (QSAoI), a novel metric that rigorously captures the trade-offs among freshness and semantic efficiency of high-level features in real-time communication in the FBL regime.
Guided by this metric, we propose a novel foundation model-based efficient co-designed framework to minimize the expected QSAoI over wireless fading channels in latency-constrained semantic communication.
Specifically, we formulate a non-linear joint optimization problem to dynamically optimize the block-wise mixed-precision quantization (MPQ) strategy and the physical blocklength.
To efficiently resolve this complex problem, we develop a high-efficiency low-complexity algorithm based on fixpoint inspection and bisection search.
Extensive simulations validate that our proposed algorithm dynamically adapts the semantic quantization precision to varying channel conditions, effectively minimizing the expected QSAoI compared to baselines.
\end{abstract}

\vspace{-.2cm}

\vspace{-.1cm}
\section{Introduction}
Towards the escalating demand for massive connections and high-speed communication, the advent of the 6G era has catalyzed a paradigm shift from conventional bit-level wireless communication towards content-aware or task-oriented cognitive networks.
In this context, semantic communication has emerged as a prospective technology \cite{chaccour2024less, li2025cognitive}.
An intelligent semantic-aware communication system aims to extract the meaningful content first, thus directly reducing redundancy under competitive resource management. 

The Age of Information (AoI) has been extensively investigated to characterize the freshness of information updates in latency-critical applications.
However, traditional AoI uniformly penalizes information without considering the actual content or semantic significance of the delivered messages with direct task accuracy oriented goals\cite{sun2017update,yu2020average}.
Despite their theoretical elegance, the majority of existing goal-oriented metrics are confined to idealized upper-layer abstractions.
To address the latency-critical scenarios, 
FBL \cite{fbl_1} information theory has been introduced to more accurately characterize the inevitable reliability degradation in short-packet communications.
While recent works have extensively researched the time-oriented AoI performance under the FBL regime\cite{zhang2025information}, 
such literature isolates the temporal freshness from the actual semantic-level significance.
The performance analysis of goal-oriented semantic metrics in the FBL regime remains in its infancy. 
Moreover, existing studies overwhelmingly isolate physical blocklength allocation from semantic precision and semantic computing delay\cite{meng2023toward}, completely ignoring the dynamic interplay between FBL-induced packet loss and computing-induced semantic penalty. 

Considering end-to-end AoI, 
deploying large AI models at the network edge for semantic feature extraction inevitably incurs substantial computing latency.
To deal with computing latency and transmission latency, model compression techniques\cite{quant_1} have become indispensable to enable real-time semantic communication. 
Specifically, dynamic mixed-precision quantization allows the system to flexibly compress high-precision semantic information into discrete packets of variable sizes\cite{zhang2025latency}. 
However, existing compression schemes are predominantly designed for inference and are rarely co-optimized with physical-layer transmission mechanisms\cite{park2025vision}.
More critically, even when adaptive schemes are considered, existing studies overlook the necessity of incorporating the dynamic quantization process, along with its resulting semantic accuracy degradation and end-to-end time efficiency, into the system's objective function.

Motivated by these challenges, we propose a semantic communication system co-designed for semantic compression and adaptive resource allocation. 
By leveraging a pretrained foundation model, we focus on the dynamic block-wise mixed-precision quantization (MPQ) strategy at the computing stage.
Furthermore, to ensure continuous and fresh semantic updates, we develop a co-designed resource scheduler that jointly optimizes the semantic quantization precision and the physical blocklength in FBL regime.
The contributions of this work can be summarized as follows:
\begin{itemize}
    \item {\bf Split-inference semantic framework with block-wise adaptive MPQ}: 
    We propose a semantic cross-layer architecture that seamlessly integrates the zero-shot generalization of the foundation model \cite{awais2025foundation} with edge computing.
    By introducing a block-wise MPQ strategy, our framework dynamically compresses high-dimensional semantic features, thereby enabling significant adaptation to wireless channel conditions without the prohibitive overhead of domain-specific retraining.

    \item {\bf Novel QSAoI metric in semantic communications}: To capture the effective freshness of semantic information in latency-constrained short-packet semantic communication, we propose a novel quantized semantic Age of Information (QSAoI). 
    This metric analytically couples information freshness with semantic fidelity under FBL constraints.
    
    \item {\bf Low-complexity co-designed QSAoI minimization}: To solve the complex QSAoI minimization problem, we develop a highly efficient algorithm that jointly determines the quantization precision and physical resource allocation utilizing fixpoint inspection.
    Extensive simulations demonstrate that our strategy significantly outperforms baselines by executing a dynamic semantic-level transition, scaling from basic semantics at low SNRs to high-resolution semantics under favorable channels.
\end{itemize}

The remaining sections are organized as follows. 
In Section~\ref{sec_framework}, we state the system framework. 
Section~\ref{sec_algorithm} presents the proposed feature compression strategy and the QSAoI minimization algorithm, followed by extensive performance evaluations in Section~\ref{sec_results} and conclusion in Section~\ref{sec_conclusion}.

\section{Semantic Communication Framework}
\label{sec_framework}
\vspace{-.1cm}

\begin{figure}[!t]
	\centering
	\includegraphics[width=0.46\textwidth]{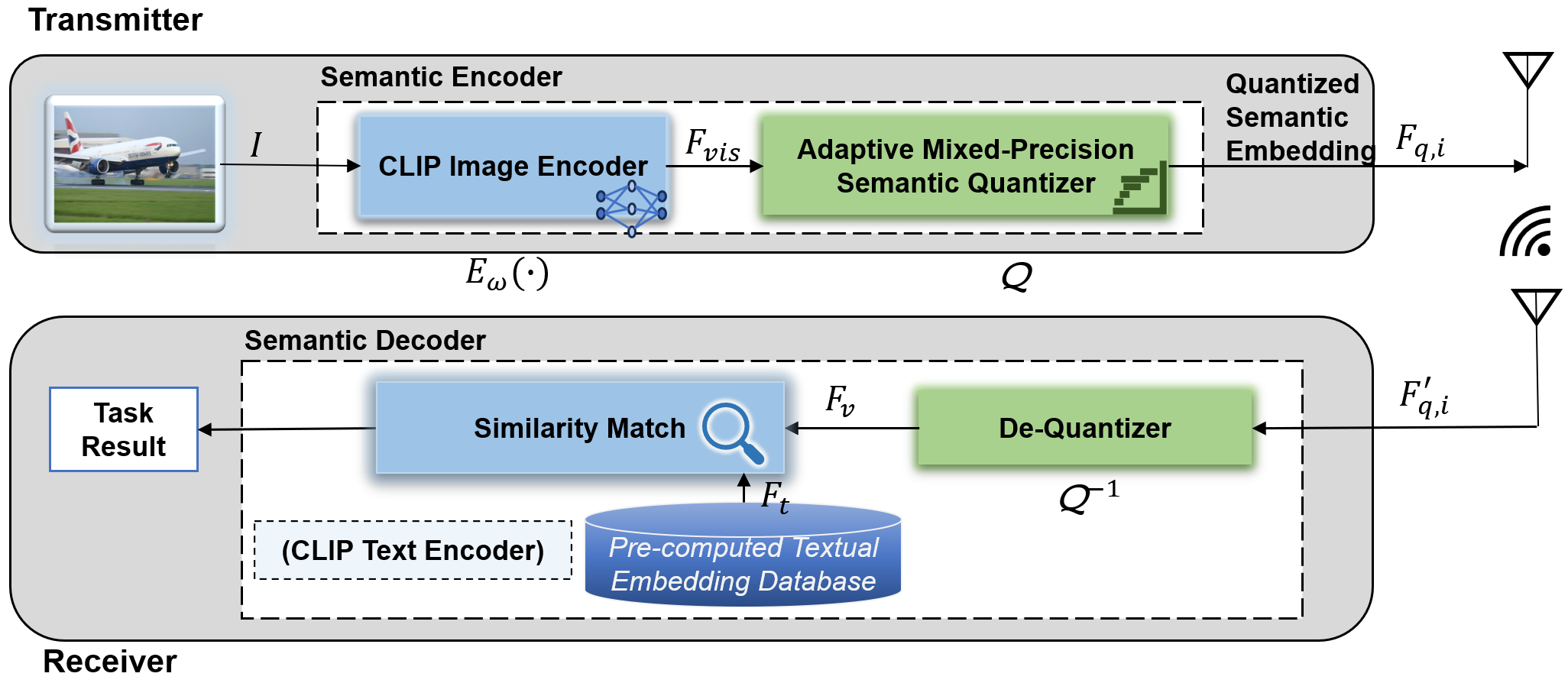}
	\caption{System architecture for foundation model-based semantic communications.}\label{fig_fra}
    \vspace{-0.7cm}
\end{figure}

In this work, we investigate a cognitive task-oriented semantic communication system as shown in Fig. \ref{fig_fra}, which comprises a resource-constrained edge device (transmitter) and an edge server (receiver).
Constrained by the limited computational capacities typically in mission-critical applications like smart factories, we adopt a split-inference paradigm where the edge device acts exclusively as a visual feature extractor utilizing a pre-trained Contrastive Language-Image Pretraining (CLIP) foundation model. 
Specifically, the edge device operates exclusively as a feature extractor to generate continuous semantic embeddings.
To alleviate bandwidth limitations, these embeddings are subsequently dynamically quantized and transmitted over a fading channel to the edge server, which executes the computationally and memory intensive task inference.
To rigorously orchestrate this cross-layer process, we introduce the QSAoI metric as our core objective. Driven by this metric, the system couples semantic utility precision with physical transmission reliability and delay.

\vspace{-0.1cm}
\subsection{Semantic Encoder}
\vspace{-0.1cm}
The edge device comprises a camera and a semantic encoder, which consists of a transformer-based CLIP image encoder\cite{radford2021learning}.
Specifically, a captured image $I \in \mathbb{R}^{H \times W \times C}$ is first resized and normalized, and then input to the image encoder. The output of this encoder is a feature of vision embedding, which we denote by
\vspace{-0.2cm}
\begin{equation}
F_{vis} = E_{\omega} (I), 
\end{equation}
where $F_{vis} \in \mathbb{R}^{K}$, and $E_{\omega} (\cdot )$ indicates the feature extractor network with learnable parameters $\omega$ that would be applicable to various data leveraging the inherent generalization capabilities of the foundation model.

Although this 512-dimensional 32-bit floating point (FP32) output high-level semantic feature ($1.6 \times 10^4$) is significantly smaller than the raw image, its volume remains prohibitive for latency-constrained semantic communication.
In this way, we introduce an adaptive block-wise mixed-precision semantic quantizer followed by the image encoder.
This adaptive quantizer assigns each block of the output embedding from a set of ultra-low bit choices of \{1, 2, 3\} in a hardware-friendly scalar quantization manner.
The proposed adaptive block-wise quantizer can be denoted as 
\vspace{-0.2cm}
\begin{equation}
\label{eq_quan}
F_{q,i} = \mathcal{Q}(F_{vis}), 
\end{equation}
where $\mathcal{Q}(\cdot)$ denotes quantization with non-learnable factors, and $i$ indexes the blocks in the output high-level embedding.

\vspace{-0.2cm}
\subsection{Wireless Channel Model}
\vspace{-0.1cm}
To satisfy the stringent end-to-end latency requirements, we model the wireless transmission in the FBL regime for short-packet digital communications.
Accordingly, the transmission latency is determined by $mT_s$, where the blocklength $m$ represents the number of transmitted symbols, and the symbol duration $T_s$ is reciprocal to the system bandwidth $B$. 
The computational latency is bounded by a constant $\tau_p$, comprising the semantic encoding, compression, and decoding processes.
At the start of each coherence time, a pilot yields estimated channel state information (CSI) $h$. 
The estimation error is defined as $\Tilde{h}$.
By treating the CSI estimation error as interference, the effective signal-to-noise ratio (SNR) is
\vspace{-0.2cm}
\begin{align}
\label{eq_system_gammak}
   \gamma(h) = \frac{|h|^2 P}{|\Tilde{h}|^2 P + N_0},
\end{align}
where $P$ is the transmit power and  
$N_0$ is the noise power. 
Based on this effective SNR, the determination of quantization strategy and resource scheduling is conducted. 
Furthermore, for a single transmission over wireless communication with SNR $\gamma$, blocklength $m$ and packet size $D$, the decoding error probability can be tightly denoted as \cite{fbl_1}
\vspace{-0.2cm}
\begin{equation}
\label{eq_fbl_2}
\varepsilon(m, \gamma, D) = Q\left(\sqrt{\frac{m}{V(\gamma)}} \left(C(\gamma) - \frac{D}{m}\right)\right),
\end{equation}
where 
$V(\gamma ) = (1 - \frac{1}{(1+\gamma )^{2} })(\log_{2}{e}) ^{2}  $ 
is the wireless channel dispersion,
$C(\gamma ) = \log_{2}{(1+\gamma)}$ is the Shannon capacity, and 
$Q(x) = \frac{1}{\sqrt{2\pi } } \int_{x}^{\infty } \exp(-\frac{t^{2} }{2} ) \mathrm{d}t $ represents the complementary Gaussian cumulative distribution function. 

\subsection{Semantic Decoder}
On the edge server (receiver) side, the received data $F_{q, i}'$ is dequantized to FP32 for the following computation.
The dequantization and reconstruction can be denoted as 
\vspace{-0.2cm}
\begin{equation}
F_{v} = \mathcal{Q}^{-1} (F_{q, i}'),
\end{equation}
where $\mathcal Q^{-1} (\cdot)$ denotes dequantization. 
Then $F_{v}$ is input into the similarity match, where a pre-computed textual embedding database is utilized. 
This comprehensive storage of semantic labels facilitates direct inference on unseen objects without any task-specific retraining.

\vspace{-0.1cm}
\section{Efficient Block-Wise Bit Determination and QSAoI Minimization}
\vspace{-0.1cm}
\label{sec_algorithm}

\subsection{Block-Wise MPQ Bit Determination of Semantics}
\vspace{-0.1cm}
To efficiently compress the continuous semantic embeddings extracted by the visual encoder, we introduce a block-wise mixed-precision quantization (MPQ) strategy.
We partition the high-dimensional semantic embedding into $N$ distinct blocks, denoted by the set $\{f_1, f_2, \dots, f_N\}$. 
For each block, the system dynamically assigns a specific quantization precision selected from a predefined candidate set $\mathcal{L} = \{d_1, d_2, \dots, d_J\}$. 
The ultimate objective of this design is to minimize the quantization distortion.
The system must minimize the total feature reconstruction error while strictly satisfying the dynamic payload constraint imposed by the overarching cross-layer scheduler.
Consequently, we formulate this block-wise bit allocation problem as follows:
\vspace{-0.1cm}
\begin{align}
\min_{\mathbf{x}} \quad & \sum_{i=1}^{N} \sum_{j=1}^{J} x_{i,j} \| f_i - \hat{f}_{i,j} \|_2^2 \label{eq1}\\
\text{s.t.} \quad & \sum_{j=1}^{J} x_{i,j} = 1, \quad \forall i \in \{1, 2, \dots, N\}, \tag{\ref{eq1}{a}}\\
& \sum_{i=1}^{N} \sum_{j=1}^{J} x_{i,j} d_j \le D, \tag{\ref{eq1}{b}}\\
& x_{i,j} \in \{0, 1\},  \forall i \in \{1, \dots, N\}, \forall j \in \{1, \dots, J\} \tag{\ref{eq1}{c}},
\end{align}
where $\mathbf{x}$ denotes the decision matrix collecting all binary variables $x_{i,j} \in \{0, 1\}$, $x_{i,j}=1$ indicates that the \mbox{$i$-th} semantic block $f_i$ in FP32 is quantized using the $j$-th precision $d_j$, and $\| f_i - \hat{f}_{i,j} \|_2^2$ represents the squared Euclidean distance quantifying the information loss of the $i$-th block under precision $d_j$.  
The first constraint ensures that exactly one precision level is assigned to each individual block. The second constraint fundamentally guarantees that the total aggregated bits across all blocks do not exceed the discrete payload budget $D$, which is also the package size in the problem \eqref{eq_fbl_2} and strictly dictated by the time-aware metric evaluation.

To satisfy strictly low-latency network, we pre-compute the optimal block-wise bit allocations for various bit budgets $D$ offline. These allocation profiles, alongside quantization parameters, are stored within a shared semantic knowledge base at both the transmitter and receiver. 
Ultimately, this mapping between the allocated bit budget $D$ determined by quantized bits of features and the corresponding task accuracy formally establishes the semantic utility function $A(D)$, explicitly characterizing the fundamental trade-off between semantic precision and transmission payload size.

\vspace{-0.1cm}
\subsection{Joint Optimization for QSAoI}
\vspace{-0.1cm}
The temporal metrics fail to recognize that rapidly delivered features remain invalid if their semantic utility significantly degrades the task inference accuracy.
To rigorously orchestrate the communication and computing resources in semantic communication, we formally introduce the QSAoI metric, which jointly evaluates latency and semantic utility.

The total semantic update latency comprises the physical transmission delay $mT_s$ and the processing overhead $\tau_p$.
Over a Rayleigh fading channel, the average decoding success probability in the FBL regime is given by 
$\bar{S}_p (m)= 1 - \int_h \varepsilon(m, \gamma(h), D)p(h)dh$, where $p(h)$ is the probability density function of $h$.
By using the semantic utility function $A(D)$ defined in the last subsection, we orchestrate the cross-layer parameters to formulate the effective semantic success probability as $S_\text{eff} = A(D)\bar{S}_p (m)$.
Under a zero-wait status generation policy, the number of transmission attempts required to achieve a semantically successful update follows a geometric distribution with mean $1/S_\text{eff}$. The expected inter-update time is consequently $\frac{mT_s + \tau_{p}}{S_\text{eff}}$. 
By geometrically evaluating the area under the sawtooth age trajectory \cite{yu2020average}, the expected QSAoI is derived as $\frac{1}{2}(mT_s + \tau_{p}) + \frac{mT_s + \tau_{p}}{S_\text{eff}}$. 
Consequently, the joint optimization problem to minimize the expected QSAoI is formulated as follows:
\vspace{-0.1cm}
\begin{align}
    \min_{m, D} \quad &\frac{1}{2} ( mT_s + \tau_{p} ) + \frac{mT_s + \tau_{p}}{A(D) \left( 1 - \int_{h} \varepsilon(m, \gamma(h), D) p(h) dh \right)}, \label{eq2} \\
    \text{s.t.} \quad 
    & m > 0, \tag{\ref{eq2}{a}} \\
    & D \in \mathcal{D}, \tag{\ref{eq2}{b}} 
\end{align}
where $\mathcal{D} = \{D_1, D_2, \dots, D_G\}$ denotes the finite set of possible semantic payload sizes. 
This overall objective function reveals that degraded semantic precision or severe channel impairments will significantly amplify the expected age penalty.
However, this problem is non-convex due to the discrete feasible set $\mathcal{D}$ and non-linear fractional objective involving an integral of the $Q$-function.

To tackle this non-convex problem, we develop an efficient iterative algorithm based on bisection search.
By introducing a continuous auxiliary variable $\mu$ to represent the objective value, we isolate the physical blocklength variable $m$ and mathematically transform the objective equality to explicitly reveal the dynamic boundary condition:
\vspace{-0.15cm}
\begin{align}
\label{eq_m}
    m = g(m, \mu, D) \triangleq \frac{1}{T_s} \left[ \frac{\mu}{ \frac{1}{2} + \frac{1}{A(D)\bar{S}_p (m)}  } -\tau_p       \right] .
\end{align}

\begin{algorithm}[t]
\small
\caption{\textbf{Algorithm for Efficient QSAoI Minimization}}
\label{algorithm1}
\begin{algorithmic}
    \STATE \textbf{Initialize:} the lower bound $\mu^{(1)}_{l}$ and upper bound $\mu^{(1)}_{u}$ for the optimal QSAoI, and the full set $\Phi _{0}$ for all possible payloads.
    \REPEAT
        \STATE $\mu_{th} \leftarrow (\mu_{l} + \mu_{u})/2$, and let $\Phi_{1} \leftarrow \Phi_{0}$
        \FOR{each $D \in \Phi_0$}
            \IF{the trial threshold $\mu_{th}$ is infeasible (i.e., $\nexists m > 0$ s.t. $m = g(m, \mu_{th}, D)$)}
                \STATE Remove $D$ from $\Phi_{1}$
            \ENDIF
        \ENDFOR
        \IF{$\Phi_{1}$ contains more than one element}
            \STATE $\mu_{u} \leftarrow \mu_{th}$, $\Phi_0 \leftarrow \Phi_{1}$ 
        \ELSIF{$\Phi_{1}$ is empty}
            \STATE $\mu_{l} \leftarrow \mu_{th}$ 
        \ELSE
            \STATE \textbf{Break loop} \quad \textit{\% Exactly one optimal candidate remains}
        \ENDIF
    \UNTIL{convergence}
    \STATE \textbf{Output:} Optimal $D^* \in \Phi_{1}$ and blocklength $m^*$ satisfying $m^* = g(m^*, \mu_{th}, D^*)$.
\end{algorithmic}
\end{algorithm}
\vspace{-0.1cm}

We propose Algorithm \ref{algorithm1} to efficiently solve this problem via integrating an outer bisection search with an inner fixpoint inspection. 
The system initializes the boundaries $\mu_l$ and $\mu_u$, alongside a candidate set of $\mathcal{D}$ denoted by $\Phi_0$. 
By evaluating the trial threshold $\mu_{th}$ in each loop, the algorithm systematically prunes $\mu_{th}$-unachievable candidates $D$. 
Since $m$ is deeply embedded within the complex integral of error probability, we deploy an inner fixpoint inspection~\cite{yuan2025optimal}.
For every candidate $D$ in the active set $\Phi_0$, the fixpoint check solves the non-linear equation $m = g(m, \mu_{th}, D)$ to explicitly verify its subproblem feasibility. 
Subsequently, the search bounds are dynamically updated based on the number of surviving candidates in the updated set $\Phi_1$.
If multiple feasible $D$ remain, the system tightens the upper bound by setting $\mu_u = \mu_{th}$ to further press the minimization potential. 
Conversely, an empty active set prompts the system to conservatively relax the lower bound by setting $\mu_l = \mu_{th}$.
This rigorous elimination process systematically converges until the unique optimal $D^*$ and its corresponding optimal blocklength $m^*$ are jointly identified.

Let $\epsilon$ denote the tolerance for the objective value. 
The outer bisection loop requires exactly $\mathcal{O}(\log_2((\mu_u - \mu_l)/\epsilon))$ iterations to converge. 
Within each search step, the system evaluates at most $G$ candidate quantization strategies. 
For each specific candidate, the inner fixpoint inspection demands $\mathcal{O}(1)$ complexity for resolving the one-dimensional variable $m$, according to \cite{yuan2025optimal}.
Consequently, the overall computational complexity is strictly bounded by $\mathcal{O}(G \log_2((\mu_u - \mu_l)/\epsilon))$. Given the highly limited dimension of the discrete candidate set and the rapid convergence of the inner equation, this lightweight processing overhead perfectly aligns with the stringent ultra-low latency requirements of edge devices.

\vspace{-0.1cm}
\section{Numerical Results}
\label{sec_results}


\subsection{Simulation Details}
To rigorously validate the superiority of the proposed task-oriented semantic communication system, we select the CIFAR100 and the pretrained foundation model CLIP.
The overall evaluation metric is the QSAoI we proposed, 
which emphasizes that degraded semantic features amplify the penalty, as they are inapplicable even when delivered with ultra-low latency.
Based on pretrained CLIP, 
the edge server pre-computes the textual category embeddings to facilitate swift online similarity matching.
Subsequently, an offline calibration process evaluates the visual embeddings on a small calibration set (10\%) for quantization.
Then, the system systematically solves the problem \eqref{eq1} to determine and cache the optimal block-wise adaptive MPQ strategies across various discrete payload budgets.
We reasonably set the computing process time $\tau_p$ as 5 ms.
During the online execution phase, the dynamically quantized semantic payload is transmitted over a Rayleigh fading channel with an available bandwidth $B$ of 100kHz, and an SNR systematically ranging from -10 to 20 dB. 
This limited bandwidth is adopted to emulate a typical subchannel allocated to an individual edge device in a massive connected smart factory. 

\vspace{-0.2cm}
\subsection{Result Analysis}

\begin{figure}[!t]
	\centering
	\includegraphics[width=0.44\textwidth]{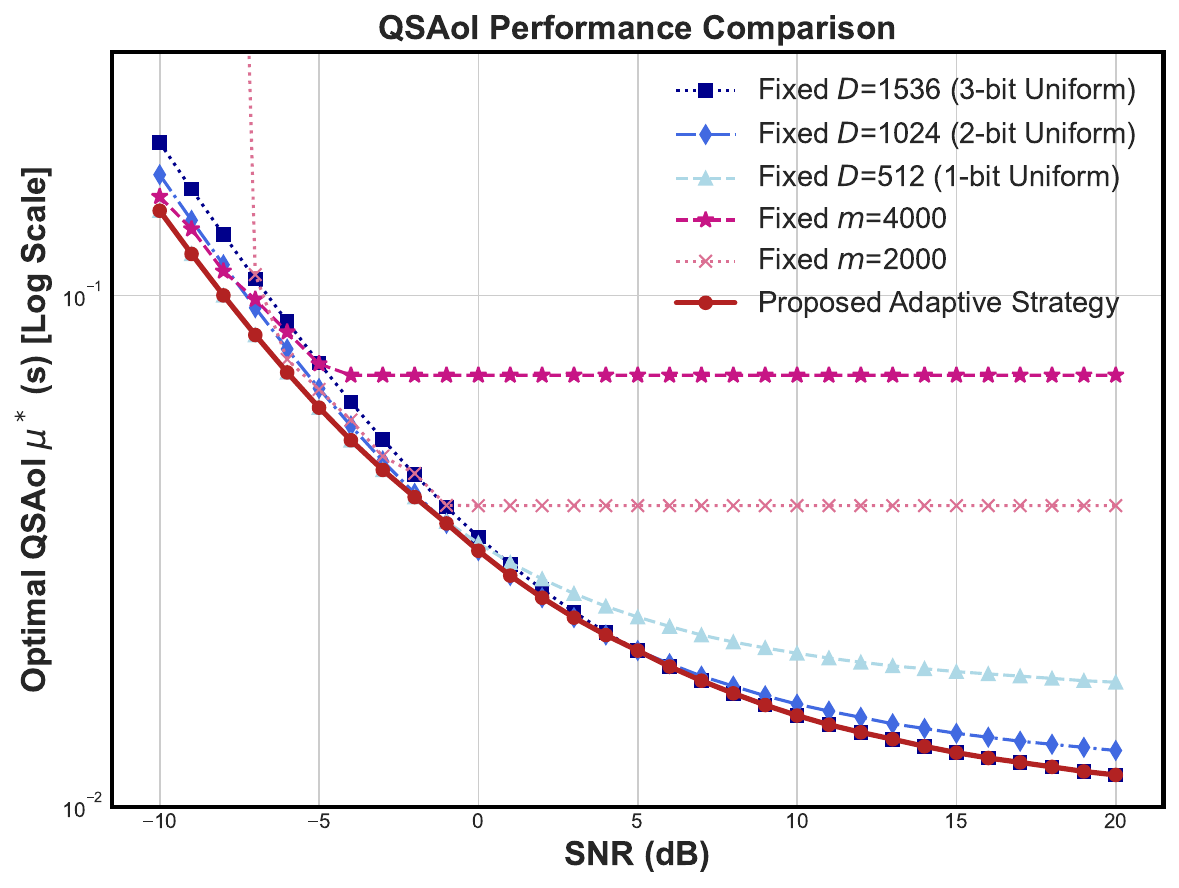}
	\caption{QSAoI comparison between the proposed adaptive strategy and uniform methods under varying SNRs.}\label{fig_1}
    \vspace{-0.5cm}
\end{figure}

Firstly, we implement the QSAoI performance comparison as illustrated in Fig. \ref{fig_1}, where the optimal QSAoI versus the SNR ratio are presented.
In order to validate the effectiveness of our proposed adaptive strategy, we compare it against 
two baselines, i.e., uniform precision (Fixed $D$) and uniform blocklength (Fixed $m$).
As expected, QSAoI decreases as channel conditions improve.
In the low-SNR range, the channel necessitates conservative semantic payloads.
The aggressive configurations, such as the 3-bit baseline (Fixed $D=1536$) and the fixed blocklength design (Fixed $m=2000$), experience severe performance degradation.
In the high-SNR range, sufficient channel capacity allows for transmitting richer features.
In stark contrast, fixed-precision baselines stagnate at a high QSAoI floor, while fixed-blocklength strategies inevitably suffer from extreme latency waste.
Specifically, our proposed method outperforms other baselines across the entire spectrum, presenting intelligent and flexible adjustments on the discrete quantization bits and the physical blocklength according to the instantaneous channel variations.

\begin{figure}[!t]
	\centering
	\includegraphics[width=0.44\textwidth]{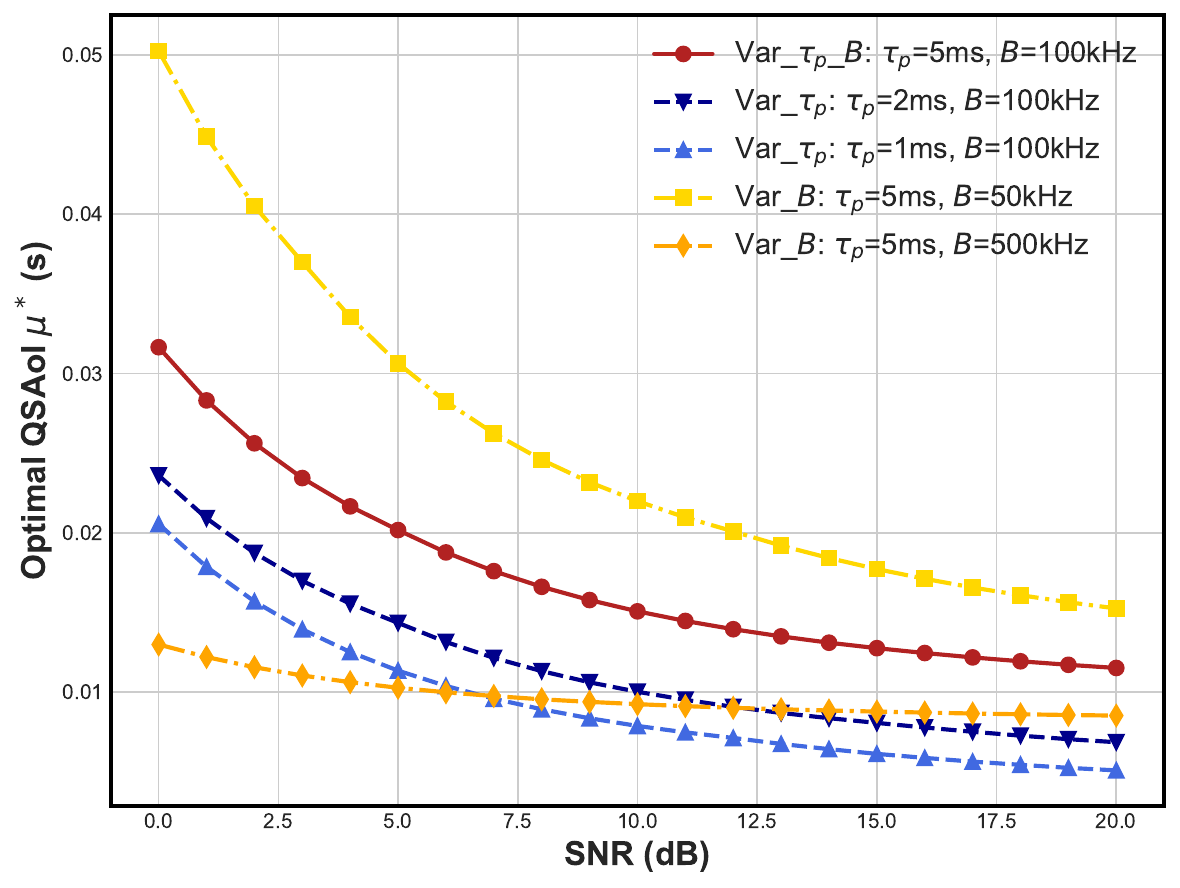}
	\caption{Comparison of QSAoI versus SNR under varying process time $\tau_p$ and bandwidth $B$.}\label{fig_2}
    \vspace{-0.4cm}
\end{figure}

Then, we consider the influence of varying process time $\tau_p$ and bandwidth $B$.
As demonstrated in Fig. \ref{fig_2}, reducing the edge computing latency $\tau_p$ from 5 ms to 1 ms consistently lowers the QSAoI across the entire channel spectrum, thereby 
highlighting the importance of reducing processing latency. 
Moreover, 
under a constrained bandwidth of 50kHz, the optimal QSAoI is sensitive to the SNR, deteriorating in the low SNR regime.
Conversely, when the system is allocated an abundant bandwidth of 500 kHz, the performance curve becomes remarkably flat and highly robust against channel fluctuations.
This physical phenomenon indicates that sufficient physical resources effectively eliminate the communication bottleneck, causing the overall system to be predominantly bounded by the inherent $\tau_p$.

\begin{figure}[!t]
	\centering
	\includegraphics[width=0.44\textwidth]{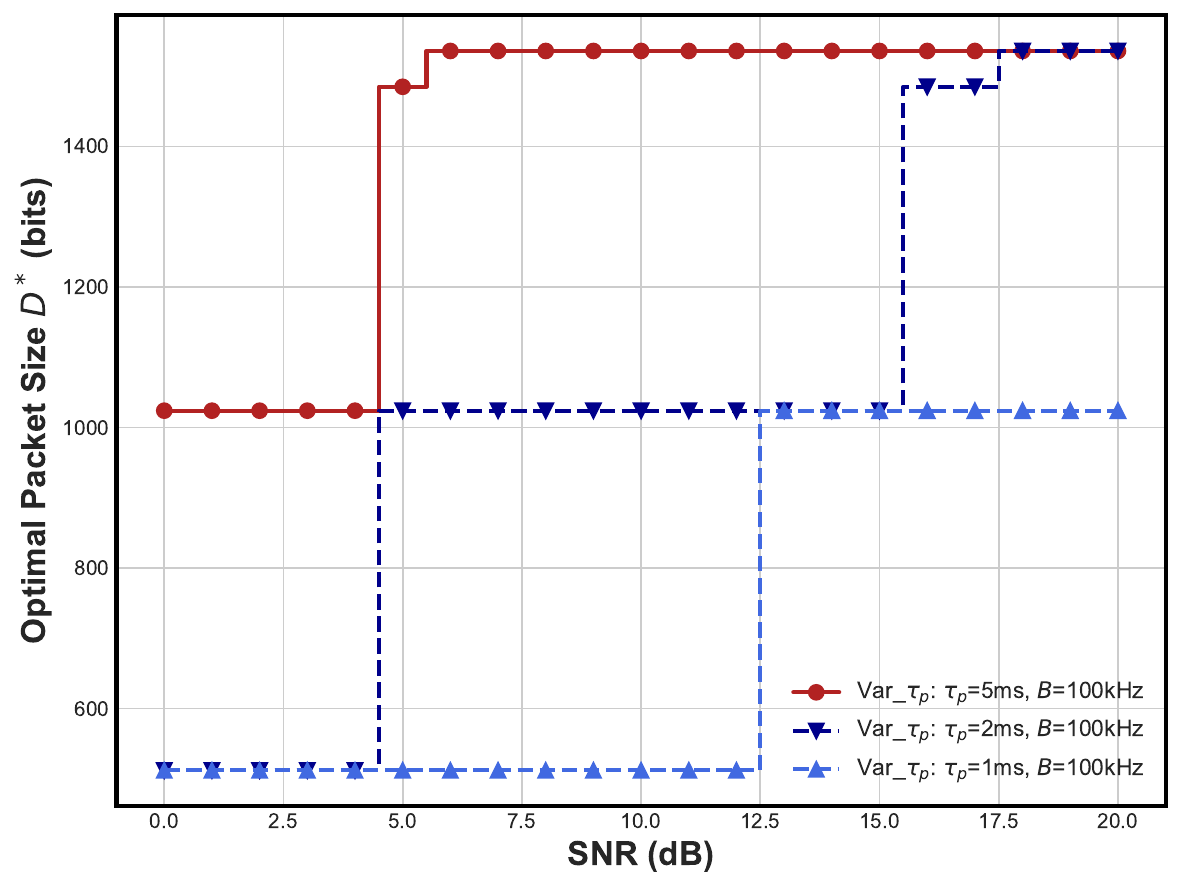}
	\caption{Optimal payload $D^*$ versus SNR under varying process time $\tau_p$ and bandwidth $B$.}\label{fig_3}
    \vspace{-0.6cm}
\end{figure}

To unveil the underlying mechanism of the proposed joint scheduling algorithm, Fig. \ref{fig_3} explicitly visualizes the optimally allocated discrete packet size $D^*$ versus SNR. 
As expected, the optimal payload exhibits discrete step transitions, reflecting the predefined mixed-precision candidate set governed by our formulation.
Crucially, this visualization reveals an intelligent scheduling mechanism across the semantic layer and physical layer regarding the computing latency $\tau_p$. 
When the terminal computing delay is dominant, the system aggressively selects a significantly larger semantic payload.
Conversely, when the edge server possesses powerful computational capabilities with low $\tau_p$, the dynamic algorithm consistently favors much smaller payloads across a significantly wider channel spectrum. 
This alternative strategy rigorously minimizes the physical transmission delay to maintain an ultra-fast semantic update cycle. 
Ultimately, this adaptive quantization trajectory proves that our algorithm effectively forces the system to dynamically shift its optimal QSAoI design according to the heterogeneous processing capabilities.

\section{Conclusion}
\label{sec_conclusion}
This paper investigated a time-aware co-designed semantic communication system.
We introduced the novel QSAoI metric, which comprehensively considers the computing and communication delay, and the semantic task accuracy. 
To deliver continuous and fresh semantic updates, we introduced a dynamic block-wise mixed-precision quantization strategy employed on the edge device, flexibly adjusting the discrete semantic payload.
Moreover, we developed an efficient joint optimization algorithm to adaptively determine the semantic quantization precision and physical resource scheduling.
Extensive empirical simulations validated that our proposed cross-layer design is significantly effective and efficient. 
This rigorous analytical framework establishes a robust theoretical foundation for future extensions into complex multi-user semantic networks.

\vspace{-0.1cm}
\bibliographystyle{IEEEtran}
\bibliography{literature}

\end{document}